\documentclass[nopubldata]{lpor2014}
\usepackage{blindtext}
\usepackage{hyperref}
\hypersetup{%
  colorlinks,
  plainpages=false,
  pdfpagelabels,
  breaklinks,
  pdfview=Fit,
  bookmarksopen,
  bookmarksnumbered,
  linkcolor=black,
  anchorcolor=black,
  citecolor=black,
  filecolor=black,
  urlcolor=LPRred
  }%
\graphicspath{{./figs/}}
\setpages[]{}
\setvolume[0]{0}

\begin{document}

\abstract{%
  Beam steering is one of the main challenges in energy-efficient and high-speed infrared light communication. To date, active beam-steering schemes based on a spatial light modulator (SLM) or micro-electrical mechanical system (MEMS) mirror, as well as the passive ones based on diffractive gratings, have been demonstrated for infrared light communication. Here, for the first time to our knowledge, an infrared beam is steered by $35^\circ$ on one side empowered by a passively field-programmable metasurface. By combining the centralized control of wavelength and polarization, a remote passive metasurface can steer the infrared beam in a remote access point. The proposed system keeps scalability to support multiple beams, flexibility to steer the beam, high optical efficiency, simple and cheap devices on remote sides, and centralized control (low maintenance cost), while it avoids disadvantages such as grating loss, a small coverage area, and a bulky size. Based on the proposed beam-steering technology, we also demonstrated a proof-of-concept experiment system with a data rate of 20 Gbps.
  }

\title{A 20-Gbps Beam-steered Infrared Wireless Link Enabled by a Passively Field-programmable Metasurface}

\author{
	Jianou Huang\inst{1,+}, 
	Chao Li\inst{1,+}, 
	Yu Lei\inst{1}, 
	Ling Yang\inst{2}, 
	Yuanjiang Xiang\inst{3}, 
	Alberto G. Curto\inst{1}, 
	Lei Guo\inst{4,*}, 
	Zizheng Cao\inst{1,*}, 
	Yue Hao\inst{2} 
	and A. M. J. Koonen\inst{1}}%

\authorrunning{J. Huang et al.}

\mail{\email{z.cao@tue.nl; guolei@cqupt.edu.cn.}}

\institute{%
Eindhoven University of Technology, Eindhoven 5600MB, The Netherlands
\and
State Key Discipline Laboratory of Wide Band-gap Semiconductor Technology, Xidian University, Xi'an 710071, China
\and
School of Physics and Electronics, Hunan University, Changsha, 410082, China
\and
School of Communication and Information Engineering, Chongqing University of Posts and Telecommunications, Chongqing, China
}


\maketitle

\section{Introduction}
\label{sec:intro}

\begin{figure*}
	\centering
	\includegraphics*[width=.6\textwidth]{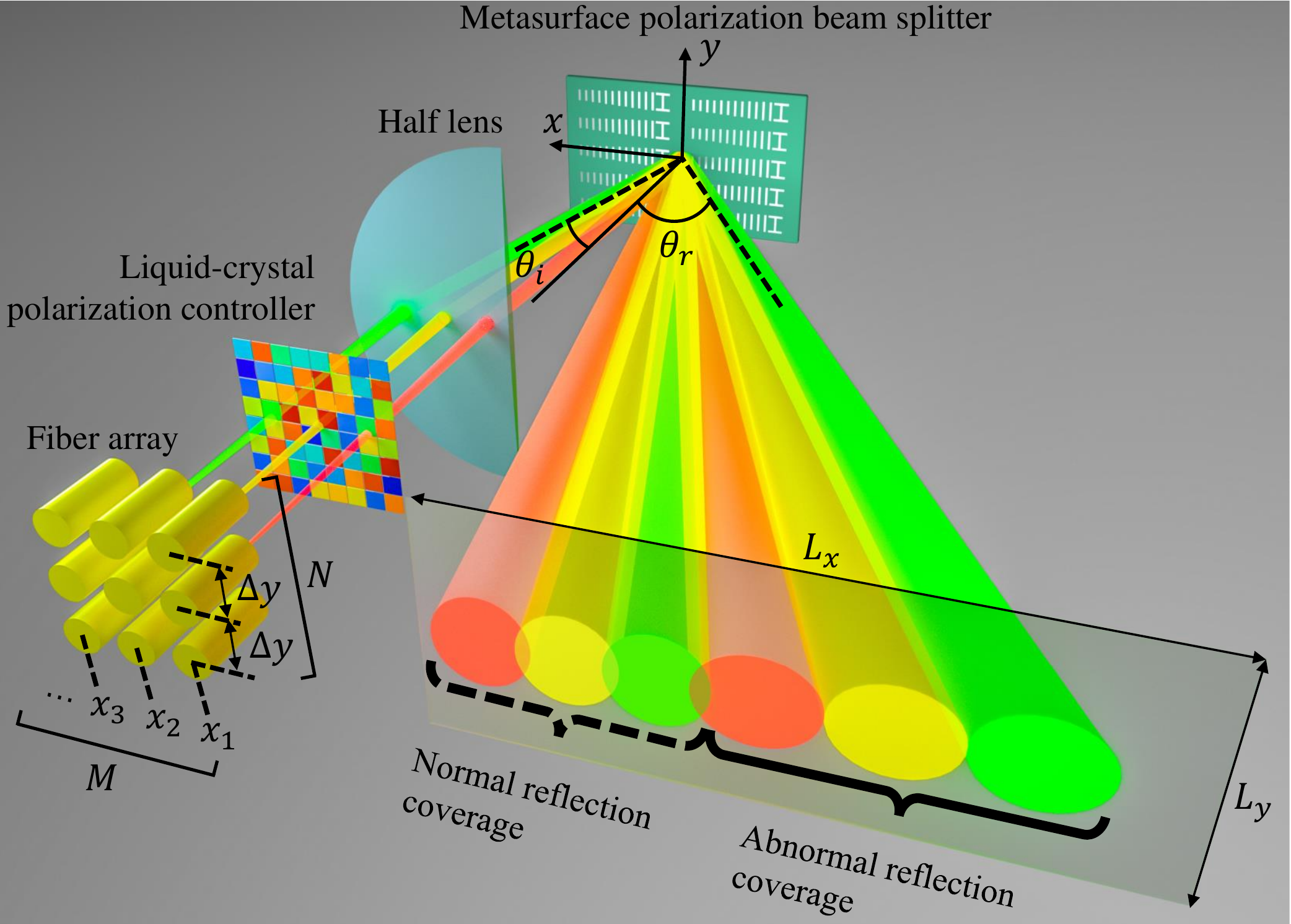}
	\caption{%
		Architecture of the proposed metasurface-based 2D IR beam-steering system.
	}
	\label{fig1}
\end{figure*}

With the explosive growth of the number of broadband mobile devices and the rapid development of the Internet of Things\cite{NctaIoT}, the booming demand for high-speed wireless connectivity is challenging the existing radio wireless communication solutions, especially in indoor scenarios. The widely used wireless technology Wi-Fi suffers from a limited data rate because the data are modulated on a narrow band and transmitted at low carrier frequencies such as 2.4 GHz and 5 GHz. More importantly, this capacity is shared among multiple users. The latest IEEE 802.11ac standard provides a channel bandwidth of up to 160 MHz in the 5 GHz spectrum, with a physical-layer data rate of up to 6.93 Gbps\cite{CiscoWiFi}. Nevertheless, the radio spectrum is still overwhelmed by the ever-increasing high-speed connection demands.

As an alternative, optical wireless communication (OWC) has recently attracted much interest because it can provide a high wireless connection capacity in an indoor environment. Hence, OWC can largely solve the shortage of radio spectrum resources\cite{koonen2017indoor}. Visible light communication (VLC)\cite{haas2015lifi} and beam-steered infrared (IR) light communication (BS-ILC)\cite{koonen2020ultra,cao2019reconfigurable} are the two main technical directions of OWC. VLC transports data over LED illumination systems and multiple users share the VLC capacity. It provides access to a bandwidth of no less than 320 THz in the 400 nm-700 nm range\cite{koonen2017indoor}. BS-ILC uses well-directed narrow IR beams to establish point-to-point communication channels. Multiple users are served by a corresponding number of IR beams. Each user has an independent connection to the transmitter, which guarantees capacity and data safety. Furthermore, directed narrow IR beams can provide high energy efficiency. Moreover, because a relatively high transmitted power of IR light is allowed in terms of eye safety, BS-ILC can achieve a very high data rate and system capacity. For VLC and BS-ILC, capacities $>$10 Gbps\cite{chun2016led} and $>$400 Gbps\cite{gomez2016design} have been demonstrated in the laboratory, respectively.

However, as a prerequisite for BS-ILC, the two-dimen-sional (2D) IR beam steering approach is still the main challenge towards the practical realization of BS-ILC. Spatial light modulator (SLM)-based\cite{gomez2014beyond,gomez2016point} and micro-electrical mechanical system (MEMS)-mirror-based\cite{wang2016full} active beam steering solutions have been proposed. When using MEMS-based mirrors, a large beam steering angle can be achieved by mechanically tuning the small mirrors, but it is difficult to realize multi-beam operation because multiple steering elements are needed, resulting in complex control schemes and system configurations. For SLM-based approaches, the IR beam can be steered by electronically tuning the phase with the SLM. No mechanical movement is introduced. Hence, quick and stable steering is realized. However, SLM-based systems are relatively bulky because complicated angle magnifiers are needed as the SLM itself can only provide very limited steering angle ($\sim3^\circ$). Additionally, grating loss is introduced, and the scalability towards many beams is limited.

Passive beam-steering solutions\cite{koonen2016ultra,chan20082,wang2018experimental,van2011two,8115031,koonen2018high} are therefore more accessible. The beam directions are mapped to the wavelengths. By inputting many wavelengths to the system through a fiber, the corresponding beam directions can be enabled simultaneously. Gratings\cite{koonen2016ultra,chan20082}, phased arrays\cite{wang2018experimental}, grating couplers\cite{van2011two}, and arrayed waveguide grating routers (AWGRs)\cite{8115031,koonen2018high} have been applied to realize fully passive 2D IR beam-steering systems. Among them, AWGR-based beam-steering modules\cite{8115031,koonen2018high} have obvious advantages: a 2D angular range of $17^\circ \times 17^\circ$, compact size, high efficiency, full area coverage, fast steering speed, 80 independent beams, and a data rate per beam of 112 Gbps, which indicate a huge system capacity of 8.96 Tbps.

Here, we present a novel solution to IR beam steering based on a passively field-programmable metasurface. Metasurfaces are an emerging approach to manipulate electromagnetic (EM) waves. They are generally created by assembling arrays of sub-wavelength resonators to provide full control of the phase, amplitude, and polarization of EM waves\cite{yu2014flat,nemati2018tunable}. Hence, actively tunable metasurfaces are widely considered for realizing beam steering\cite{wu2019dynamic,hashemi2016electronically,ma2019smart}. Tunable metasurfaces can be dynamically controlled via external stimuli, which are usually electrical biases, laser pulses, or heat inputs. Owing to the sub-wavelength scale of the resonators, controlling every resonator is difficult and would make the structure very complicated and difficult to manufacture, particularly in the optical range. Therefore, tunable metasurface-based beam steering is usually realized at microwave frequencies\cite{hashemi2016electronically,ma2019smart}. In the optical range, controlling every resonator is unrealistic, yielding very limited beam steering\cite{wu2019dynamic}.

In our system, a polarization beam splitter based on a passive gap-surface plasmon metasurface (GSPM) is applied for 2D IR beam steering. GSPM has the advantages of high efficiency, excellent control over the reflected or transmitted light, and a simple manufacturing technique\cite{ding2018review}. It has been used to realize numerous flat devices such as anomalous reflectors\cite{sun2012high}, focusing flat mirrors\cite{pors2013broadband}, phase modulators\cite{kim2019phase}, holograms\cite{yifat2014highly}, polarisation beam splitters\cite{pors2013gap}, metagratings\cite{pors2015plasmonic}, and orbital angular momentum generators\cite{ding2019dual,tang2018high,tan2019free}. With the GSPM-based beam splitter and a simple liquid-crystal polarization controller, polarization-controlled beam steering can be realized. Together with an AWGR-based beam-steering module, a polarization-wavelength-controlled 2D IR beam-steering system is achieved. The new system keeps all the advantages of AWGR-based beam steering\cite{8115031,koonen2018high} and additionally has polarization control capability. Hence, the coverage is greatly expanded. Furthermore, the grating loss existing in SLM can be avoided by using the metasurface, which improves the energy efficiency. Finally, a 20-Gbps beam-steering experiment is performed over 1.2-m free space.

\section{Architecture and Operation Principle}
\label{sec:architecture}

\begin{figure*}
	\centering
	\includegraphics*[width=1\textwidth]{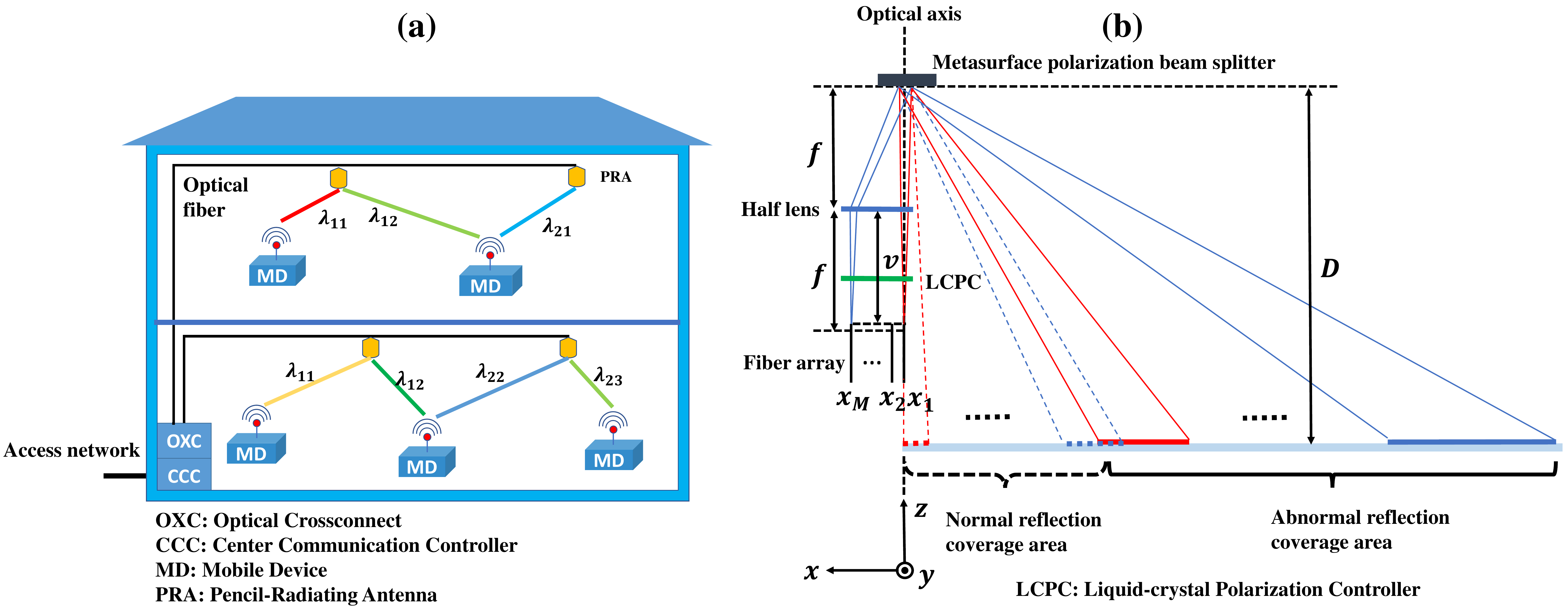}
	\caption{%
		(a) Indoor beam-steered infrared light communication network. (b) Schematic of the proposed 2D IR beam-steered infrared wireless communication system.
	}
	\label{fig2}
\end{figure*}

The proposed 2D IR beam-steering system is shown in \textbf{Figure 1}. It is composed of an AWGR-based beam-steering module\cite{8115031,koonen2018high}, a liquid-crystal polarization controller, and a metasurface polarization beam splitter. $M \times N$ outputs of the AWGR are connected to an $M \times N$ 2D fiber array. The output plane of the fiber array is placed in the object plane of a half lens. The output beams of the fiber array are first modulated by the liquid-crystal polarization controller to manipulate their polarization, following which they are deflected by the half lens. The deflected beams are then reflected by the metasurface polarization beam splitter, which is placed in the focal plane of the half lens. Both normal reflection and abnormal reflection occur on the surface of the metasurface polarization beam splitter; thus, the number of final output beams is doubled, and the total beam coverage area is greatly expanded (normal reflection coverage + abnormal reflection coverage).

The proposed beam-steering system has two tuning schemes: wavelength tuning and polarization tuning. By adjusting the input wavelength of the AWGR, one can control the position of the lighted fiber and then control the beam direction after the half lens, which further controls the emission directions of a pair of normally and abnormally reflected output beams. The polarization tuning is enabled by the liquid-crystal polarization controller and the metasurface polarization beam splitter. For a $y$-polarized incident beam, the metasurface polarization beam splitter generates a positive phase gradient along the negative x-direction; thus, the beam is abnormally reflected, as shown in Figure 1. The relation between the abnormal reflection angle $\theta_r$ and the incident angle $\theta_i$ is $\sin \theta_r = \sin \theta_i + \lambda / \Lambda$, where $\lambda$ is the wavelength and $\Lambda$ is the length of the super cell of the metasurface polarization beam splitter. The $\theta_r-\theta_i$ relation implies a critical incident angle $\theta_{ic} = \arcsin(1-\lambda / \Lambda)$. For an $x$-polarized incident beam, the metasurface polarization beam splitter acts as a mirror, and the beam is normally reflected. By adjusting the polarization of a fiber beam with the liquid-crystal polarization controller, one can control the power distribution between the corresponding normally and abnormally reflected output beams. Therefore, a wavelength-polarization-controlled 2D beam-steering approach is achieved.

In section 3, the system configuration is thoroughly discussed. The metasurface polarization beam splitter is discussed in detail in section 4.

\section{System Design}
\label{sec:sysdesign}
The concept of the indoor beam-steered infrared light communication network is shown in \textbf{Figure 2a}\cite{koonen2018high}. In each room, several pencil-beam radiating antennas can provide high-speed connections to mobile devices with well-directed narrow IR beams. To cover the whole area, 2D IR beam steering is essential. 

Figure 2b, which is the $y=0$ cross-section of Figure 1, schematically illustrates the configuration of the proposed 2D IR beam-steering system. The fiber array is placed away from the focus and closer to the lens. The distance between the output plane of the fiber array and the half lens is $v$, which defines the relative defocusing parameter $p=1-v/f$ ($0 \leq p<1$). The metasurface polarization beam splitter is placed in the focal plane of the half lens. Here, the half lens is used to avoid blocking the reflected beams.

The system aims to cover a square image area of size $L_x \times L_y$ with no spacing between adjacent beam spots, as shown in Figure 1. Since the positive phase gradient of the metasurface polarization beam splitter is along the negative $x$-direction, the wave vector in the $y$-direction follows the law of specular reflection. Therefore, each beam spot has the same width $W_y$ in the $y$-direction. To cover a length of $L_y$ at a distance $D$ from the metasurface while no spacing exists between adjacent beam spots, $W_y$ should be equal to $L_y/N$. The required lens focal length $f$ and the constant $y$-direction fiber spacing $\Delta y$ can be determined by using paraxial geometric optics\cite{8115031}:

\begin{equation}
\label{eqn:eq1}
f=\frac{L_y}{N \cdot 2 \tan\alpha} - p \cdot D
\end{equation}

\begin{equation}
\label{eqn:eq2}
\Delta y = 2f \cdot \tan\alpha \cdot  (\frac{f}{D} +p)
\end{equation}

\begin{figure*}[t]
	\centering
	\includegraphics*[width=.5\textwidth]{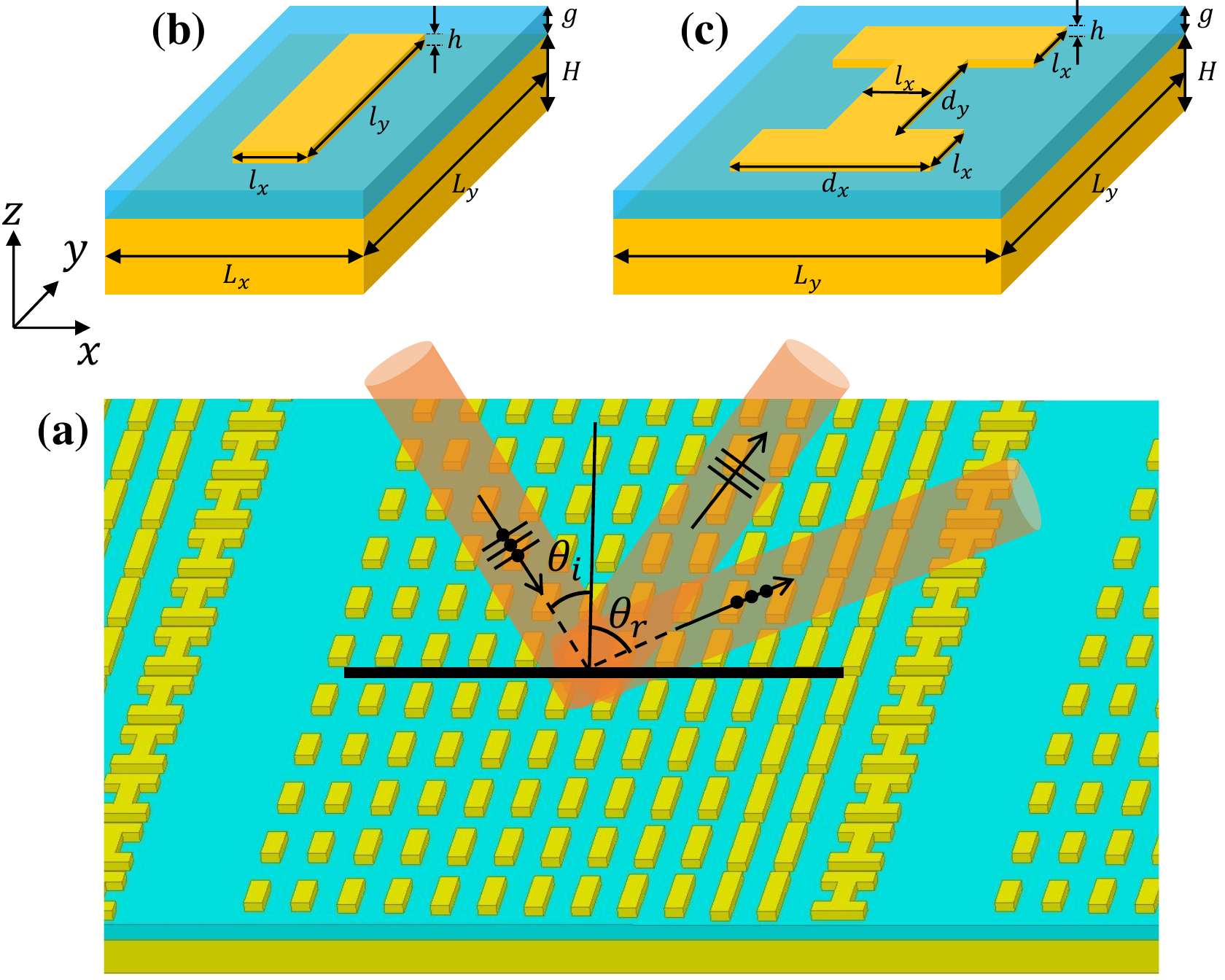}
	\caption{%
		(a) Schematic of the designed polarization beam splitter. $\theta_i$ is the incident angle, and $\theta_r$ is the abnormal reflected angle. (b) Schematic of the type 1 meta-atom. (c) Schematic of the type 2 meta-atom.
	}
	\label{fig3}
\end{figure*}

\noindent where $\tan\alpha=\lambda/(\pi w_0 )$ with the mode field radius $w_0$ of the single-mode fiber. In the normal reflection coverage area, the beam spots are all round. If the $x$-direction fiber spacing $\Delta x$ is constant and equal to $\Delta y$, the normal reflection area can be just covered. However, in this case, owing to the nonlinearity of the abnormal reflection, the abnormal-reflection coverage area cannot be fully covered. Thus, the position of each column of the fiber array $x_1, x_2...$ should be determined by making the beam spots of adjacent fiber columns tangential along the $x$-direction in the abnormal reflection coverage area:

\begin{equation}
\label{eqn:eq3}
\begin{split}
D \cdot \tan \lbrace \arcsin \lbrace \sin [ \arctan (\frac{x_{n+1} - pf \cdot \tan \alpha}{f} ) ] + \frac{\lambda}{\Lambda} \rbrace \rbrace \\
- f \cdot \tan \alpha \\
= D \cdot \tan \lbrace \arcsin \lbrace \sin [ \arctan (\frac{x_n + pf \cdot \tan \alpha}{f} ) ] + \frac{\lambda}{\Lambda} \rbrace \rbrace \\
+ f \cdot \tan \alpha
\end{split}
\end{equation}

\noindent where $n=1, 2...$, $\lambda$ is the wavelength and $\Lambda$ is the length of the super cell of the metasurface polarization beam splitter. $x_n$ ($n \geq 2$) can be obtained from the recursive formula (3) with the initial condition $x_1=0$.

A condition on $x_M$ is still needed to determine where to stop (i.e., the value of $M$). This condition can be obtained from the situation in which the normal and abnormal reflection coverage areas exactly overlap:

\begin{equation}
\label{eqn:eq4}
\begin{split}
f \cdot \tan \lbrace \arcsin \lbrace \sin [ \arctan ( -p \cdot \tan \alpha ) ] + \frac{\lambda}{\Lambda} \rbrace \rbrace \\
-\frac{2 f^2 \cdot \tan \alpha}{D} - p f \cdot \tan \alpha \\
\leq x_M < \\
f \cdot \tan \lbrace \arcsin \lbrace \sin [ \arctan ( p \cdot \tan \alpha ) ] + \frac{\lambda}{\Lambda} \rbrace \rbrace \\
 - p f \cdot \tan \alpha
\end{split}
\end{equation}

$x_M$ should take the minimum value in this range. On the other hand, $x_M$ is also limited by the critical angle of the metasurface polarization beam splitter:

\begin{equation}
\label{eqn:eq5}
x_M < f \cdot \tan(\theta_{ic}) - pf \cdot \tan \alpha
\end{equation}

\noindent where $\theta_{ic} = \arcsin (1-\lambda / \Lambda)$ is the critical angle of the metasurface polarization beam splitter. Based on the above two formulas, the following condition should be satisfied to ensure that no gap exists between the normal and abnormal reflection coverage areas:

\begin{equation}
\label{eqn:eq6}
\sin [\arctan (p \cdot \tan \alpha)] < 1 - \frac{2 \lambda}{\Lambda}
\end{equation}

Since $p \cdot \tan \alpha$ is usually small ($\sim 10^{-2}$), a clearer relation can be obtained by ignoring it:

\begin{equation}
\label{eqn:eq7}
\Lambda > 2 \lambda
\end{equation}

\noindent which is the design requirement for the metasurface polarization beam splitter. With the value range of $x_M$ and the initial condition $x_1=0$, the number of the fiber columns $M$ and the $x$-position of each column $x_1,x_2...x_M$ can be fully determined by using the recursive formula (3). Furthermore, the coverage length in the $x$-direction can now be obtained:

\begin{equation}
\label{eqn:eq8}
\begin{split}
L_x = D \cdot \tan \lbrace \arcsin \lbrace \sin [ \arctan (\frac{x_M + pf \cdot \tan \alpha}{f} ) ] + \frac{\lambda}{\Lambda} \rbrace \rbrace \\
+ f \cdot \tan \alpha
\end{split}
\end{equation}

In the abnormal reflection coverage area, no spacing or overlap exists between adjacent beam spots. In the normal reflection coverage area, no spacing or overlap exists between adjacent beam spots along the $y$-direction, while an overlap exists between adjacent beam spots along the $x$-direction because the distance between two fiber columns is less than $\Delta y$. A square image area of size $L_x \times L_y$ is fully covered.

As an example, we use the same basic parameters as Koonen et al.\cite{8115031}: $p=0.21$, $L_y=1.68$ m, $D=2.4488$ m, $N=14$, $\lambda =1.5$ $\mu$m, and $w_0=4.5$ $\mu$m. In our metasurface polarization beam splitter, $\Lambda = 4$ $\mu$m, which meets $\Lambda > 2 \lambda=3$ $\mu$m. Then, we find $f=51.2$ mm, $\Delta y=2.51$ mm, $M=9$, $L_x=2.75$ m, and $x_M=19.4$ mm. Therefore, an image area of $2.75 \times 1.68$ m$^2$ is fully covered with a $9 \times 14$ 2D fiber array (126-port AWGR), the total size of which is $19.4 \times 32.63$ mm$^2$. Compared with Koonen et al.\cite{8115031}, in which $1.68 \times 1.68$ m$^2$ is fully covered with a $14 \times 14$ 2D fiber array (196-port AWGR) of total size $32.63 \times  32.63$ mm$^2$, our system is considerably enhanced in coverage and size while using the same basic parameters.

\begin{figure}[t]
	\centering
	\includegraphics*[width=.45\textwidth]{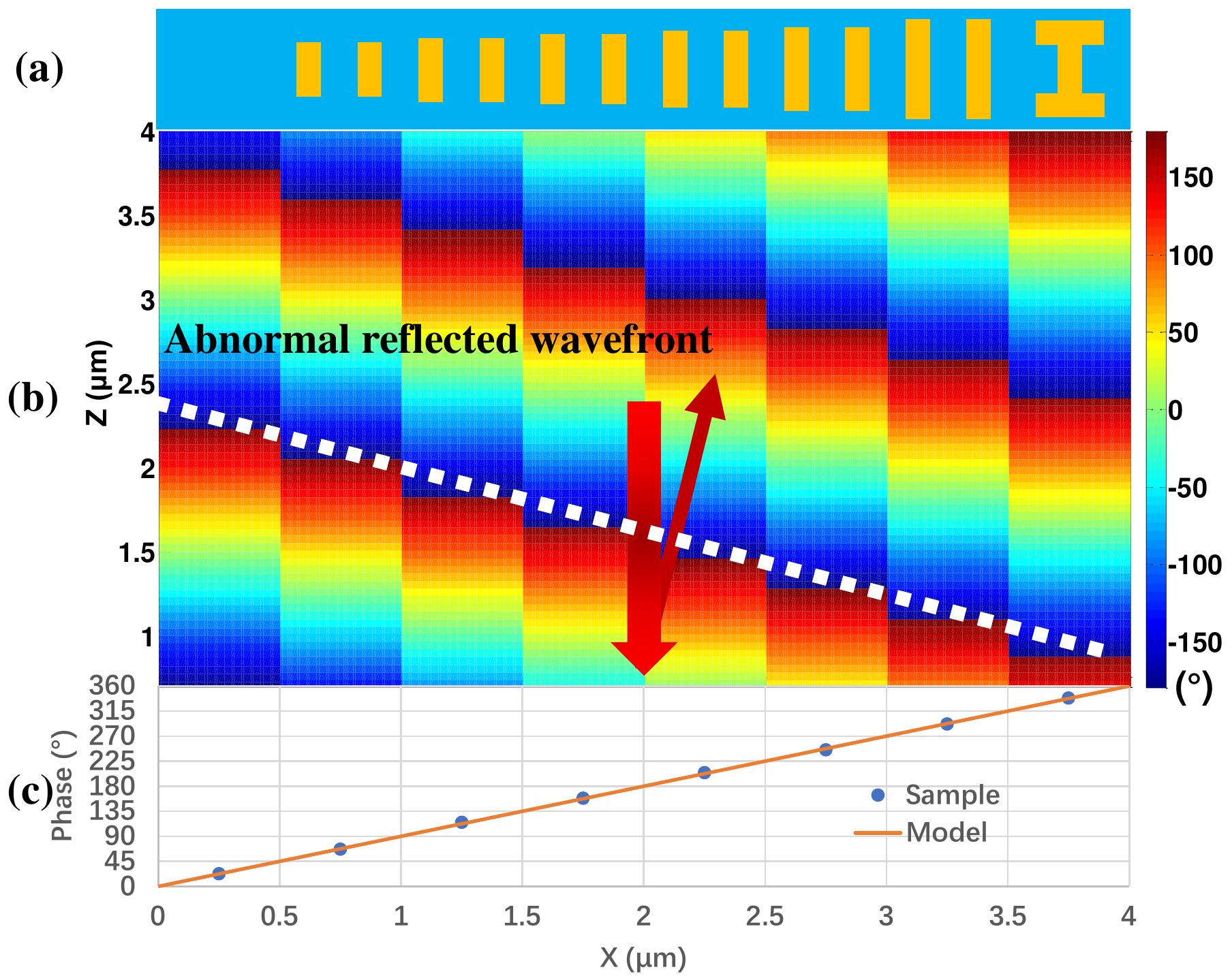}
	\caption{%
		(a) Super cell of the designed metasurface polarization beam splitter. (b) Simulated scattered $E_y$ phase patterns of all the phase units under the illumination of normally incident $y$-polarized plane wave ($\lambda=1550$ nm). (c) Scattered phase of each phase unit within a super cell.
	}
	\label{fig4}
\end{figure}

\begin{figure}[t]
	\centering
	\includegraphics*[width=.45\textwidth]{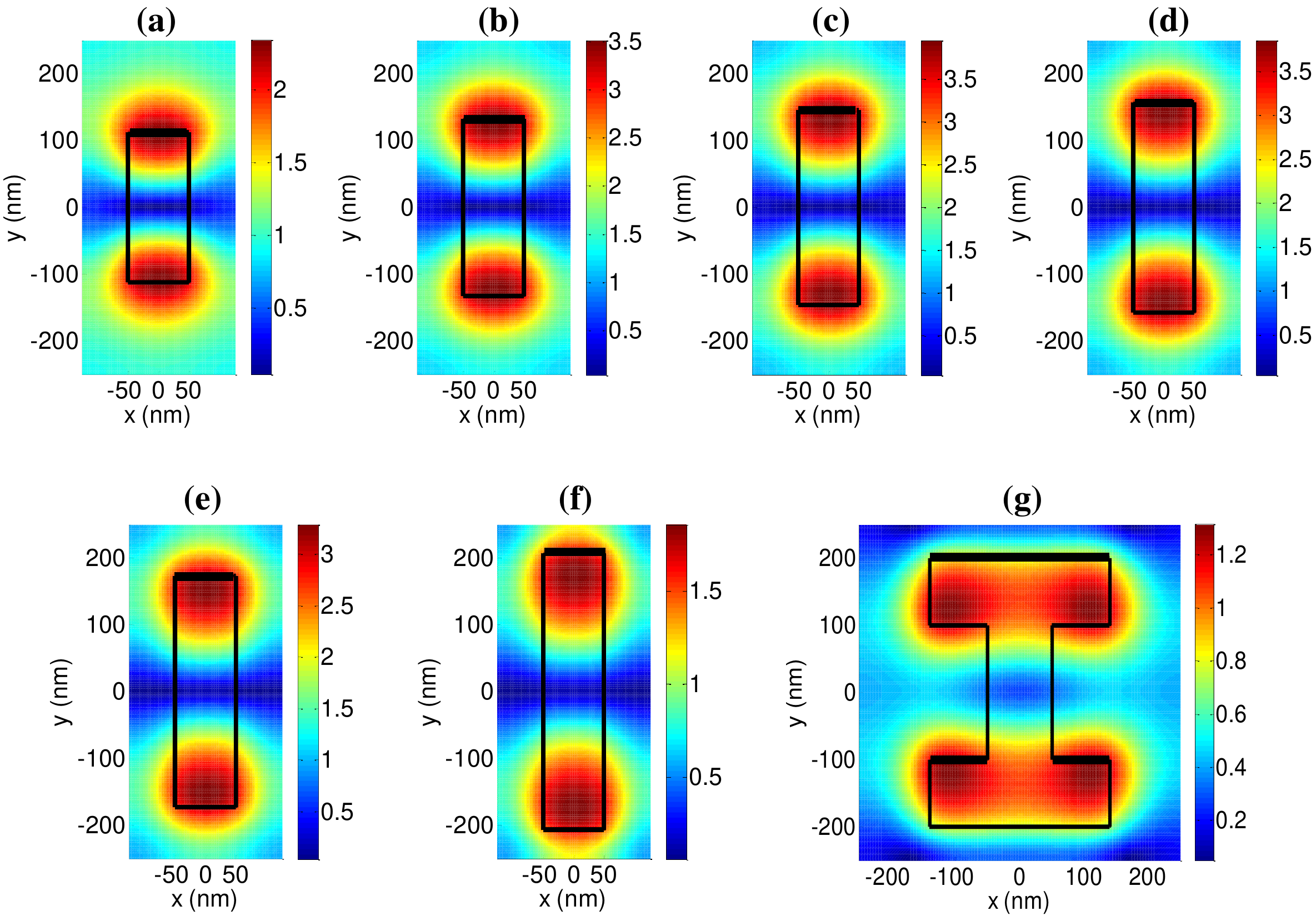}
	\caption{%
		Magnitude of the electric field at $\lambda=1550$ nm in the $xy$-plane in the center of the spacer ($y$-polarized normal incidence). (a)-(f) correspond to $l_y=225$ nm, 266 nm, 292 nm, 315 nm, 345 nm, and 413 nm type-1 meta-atoms, respectively. (g) corresponds to the type-2 meta-atom.
	}
	\label{fig5}
\end{figure}

\section{Device Design and Characterization}
\label{sec:Devicedesign}

The metasurface polarization beam splitter plays a key role in the proposed 2D IR beam-steering system. Here, to demonstrate the system concept, a GSPM-based metasurface polarization beam splitter is designed and fabricated at $\lambda=1550$ nm. \textbf{Figure 3a} schematically shows the designed polarization beam splitter, which consists of periodical arrays of gap plasmon-based meta-atoms. Two types of meta-atoms are used, type 1 and type 2, as shown in Figure 3b and Figure 3c, respectively. They are both composed of an Au ground, a SiO$_2$ spacer in the middle, and a top Au nano-pattern designed with different shapes. When the meta-atom is illuminated by an $x$-polarized or a $y$-polarized incident plane wave, electric currents are induced on both the top Au pattern and the bottom Au ground, which result in strong near-field coupling and anti-parallel electric current oscillations, forming strong magnetic resonance\cite{jung2009gap}. By varying the geometry of the meta-atoms, the reflection phase and amplitude of each unit cell can be engineered independently at the designed wavelength.

\begin{figure}[t]
	\centering
	\includegraphics*[width=.49\textwidth]{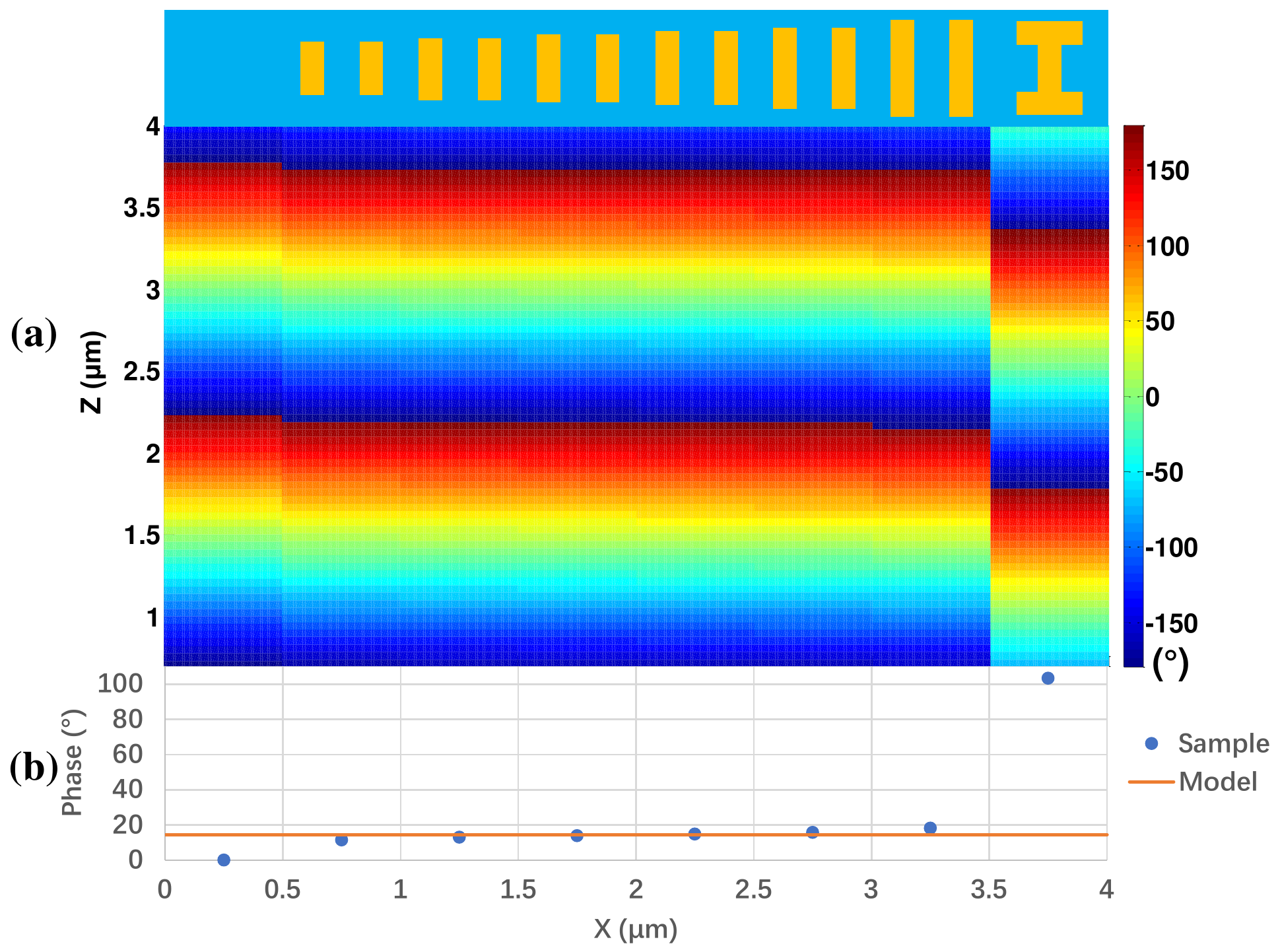}
	\caption{%
		(a) Simulated scattered $E_x$ phase patterns of all the phase units in a super cell under the illumination of normally incident $x$-polarized plane wave ($\lambda=1550$ nm). (b) Scattered phase of each phase unit within a super cell.
	}
	\label{fig6}
\end{figure}

\begin{figure}[t]
	\centering
	\includegraphics*[width=.45\textwidth]{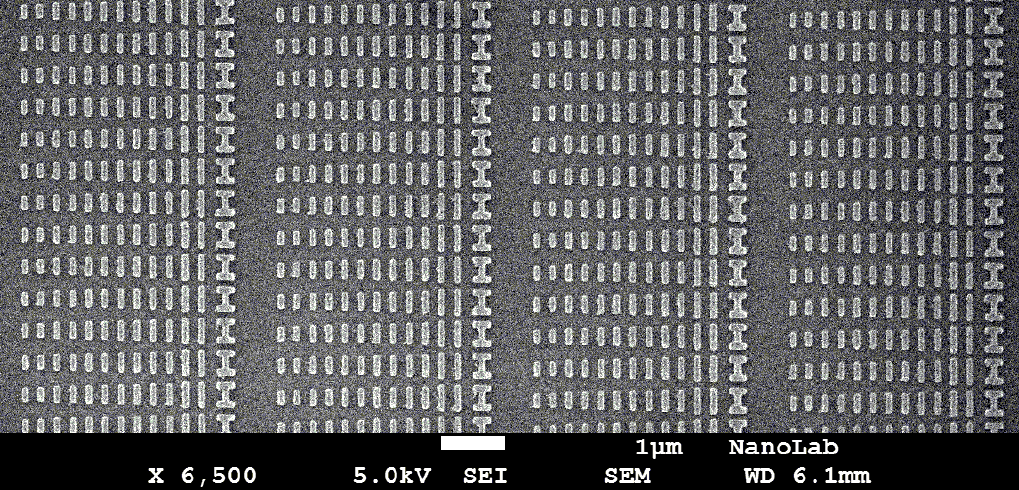}
	\caption{%
		SEM image of the fabricated metasurface polarization beam splitter.
	}
	\label{fig7}
\end{figure}

\begin{figure*}
	\centering
	\includegraphics*[width=.8\textwidth]{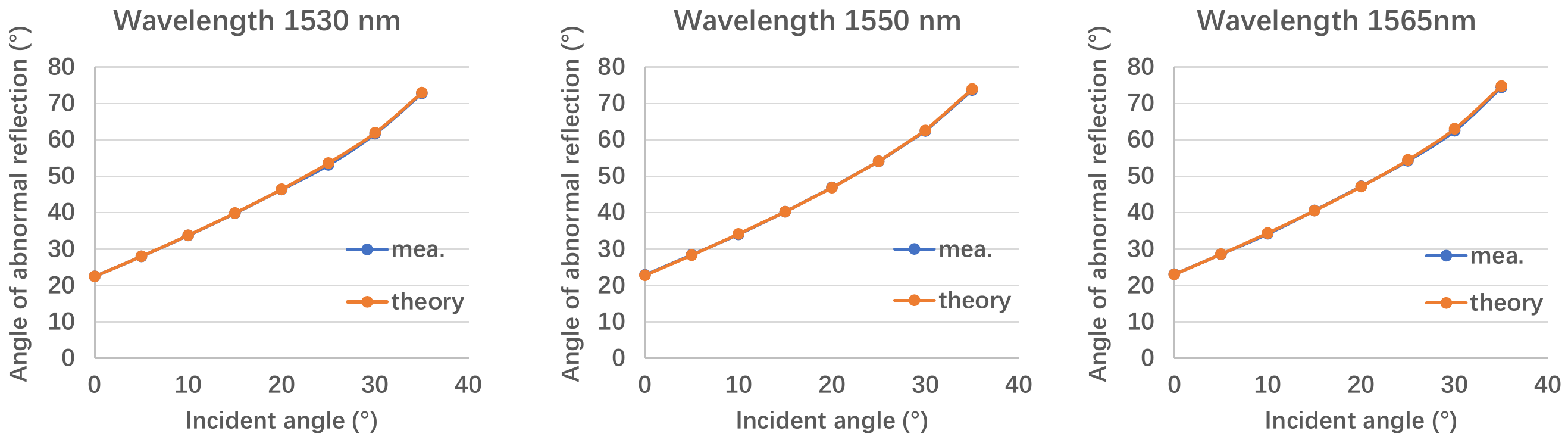}
	\caption{%
		Measured relation between the incident angle and the abnormal reflection angle.
	}
	\label{fig8}
\end{figure*}

\begin{figure*}
	\centering
	\includegraphics*[width=.85\textwidth]{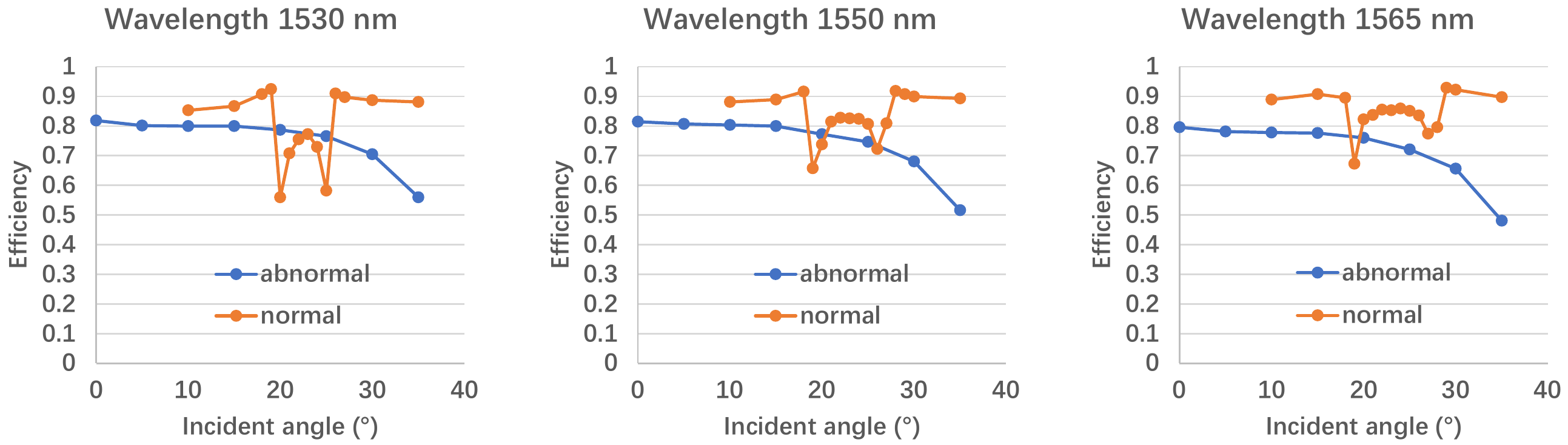}
	\caption{%
		Measured power efficiency of the abnormal reflection and normal reflection.
	}
	\label{fig9}
\end{figure*}

\begin{figure*}
	\centering
	\includegraphics*[width=.85\textwidth]{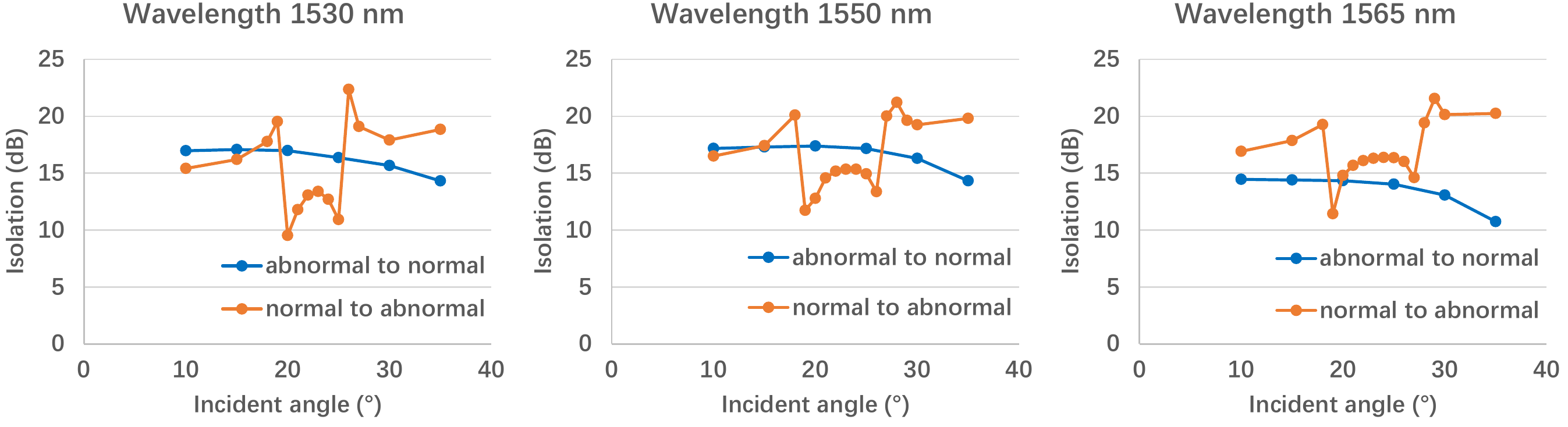}
	\caption{%
		Measured isolation between the normal and abnormal reflection.
	}
	\label{fig10}
\end{figure*}

As shown in \textbf{Figure 4a}, a super cell consists of 8 phase units is designed to cover a $2\pi$ phase with a $\pi /4$ phase interval. Each phase unit is made of meta-atoms with different geometries. In our design, all the meta-atoms share the following parameters: $L_x=250$ nm, $L_y=500$ nm, $H=200$ nm, $g=90$ nm, $l_x=100$ nm, and $h=55$ nm. The zero-phase unit has no Au pattern on the SiO$_2$ spacer. Each of the six phase units from $\pi /4$ to $6\pi /4$ is a parallel connection of 2 identical type-1 meta-atoms. The pattern lengths $l_y$ are 225 nm, 266 nm, 292 nm, 315 nm, 345 nm, and 413 nm, respectively. The $7\pi /4$ phase unit is a single type-2 meta-atom with $d_x=280$ nm and $d_y=400$ nm. All the parameters are optimized by using finite-difference time-domain (FDTD) software. Compared with the design wavelength $\lambda=1550$ nm, all the meta-atoms have subwavelength dimensions.

Figure 4b shows the simulated scattered $E_y$ phase patterns of all the phase units in a super cell under the illumination of normally incident $y$-polarized plane wave ($\lambda=1550$ nm). It can be seen that a positive phase gradient along the $x$-direction is formed, which creates an abnormal reflected wavefront. According to the generalized laws of reflection and refraction\cite{yu2011light}, for a $y$-polarized plane wave at an incident angle $\theta_i$, the abnormal reflection angle $\theta_r$ is

\begin{equation}
\label{eqn:eq9}
\sin \theta_r = \sin \theta_i + \frac{\lambda}{\Lambda}
\end{equation}

\noindent where $\Lambda$ is the length of the super cell. In our design, $\Lambda=4$ $\mu$m. It can be seen that $\theta_i$ and $\theta_r$ have a nonlinear relation, and a critical angle exists for $\theta_i$:

\begin{equation}
\label{eqn:eq10}
\theta_{ic} = \arcsin(1-\frac{\lambda}{\Lambda})
\end{equation}

\begin{figure*}
	\centering
	\includegraphics*[width=.8\textwidth]{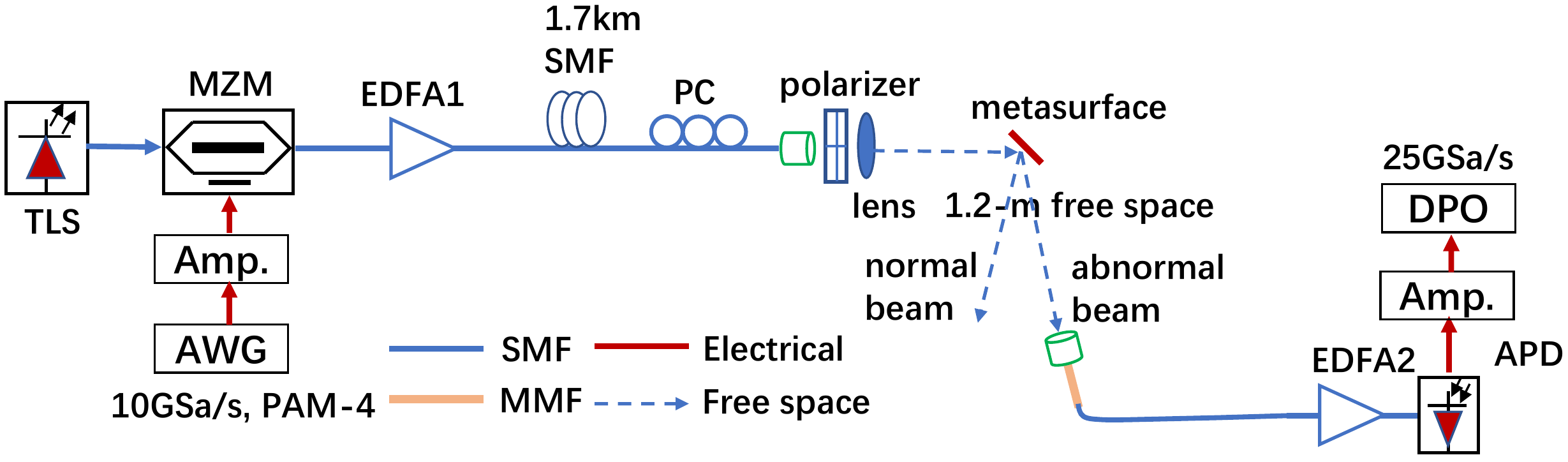}
	\caption{%
		Experimental setup of metasurface-based beam steering. TLS: tunable laser source, AWG: arbitrary waveform generator, MZM: Mach-Zehnder modulator, EDFA: erbium-doped fiber amplifier, SMF: single-mode fiber, PC: polarization controller, APD: avalanche photodiode, DPO: digital phosphor oscilloscope.
	}
	\label{fig11}
\end{figure*}

\begin{figure*}
	\centering
	\includegraphics*[width=.67\textwidth]{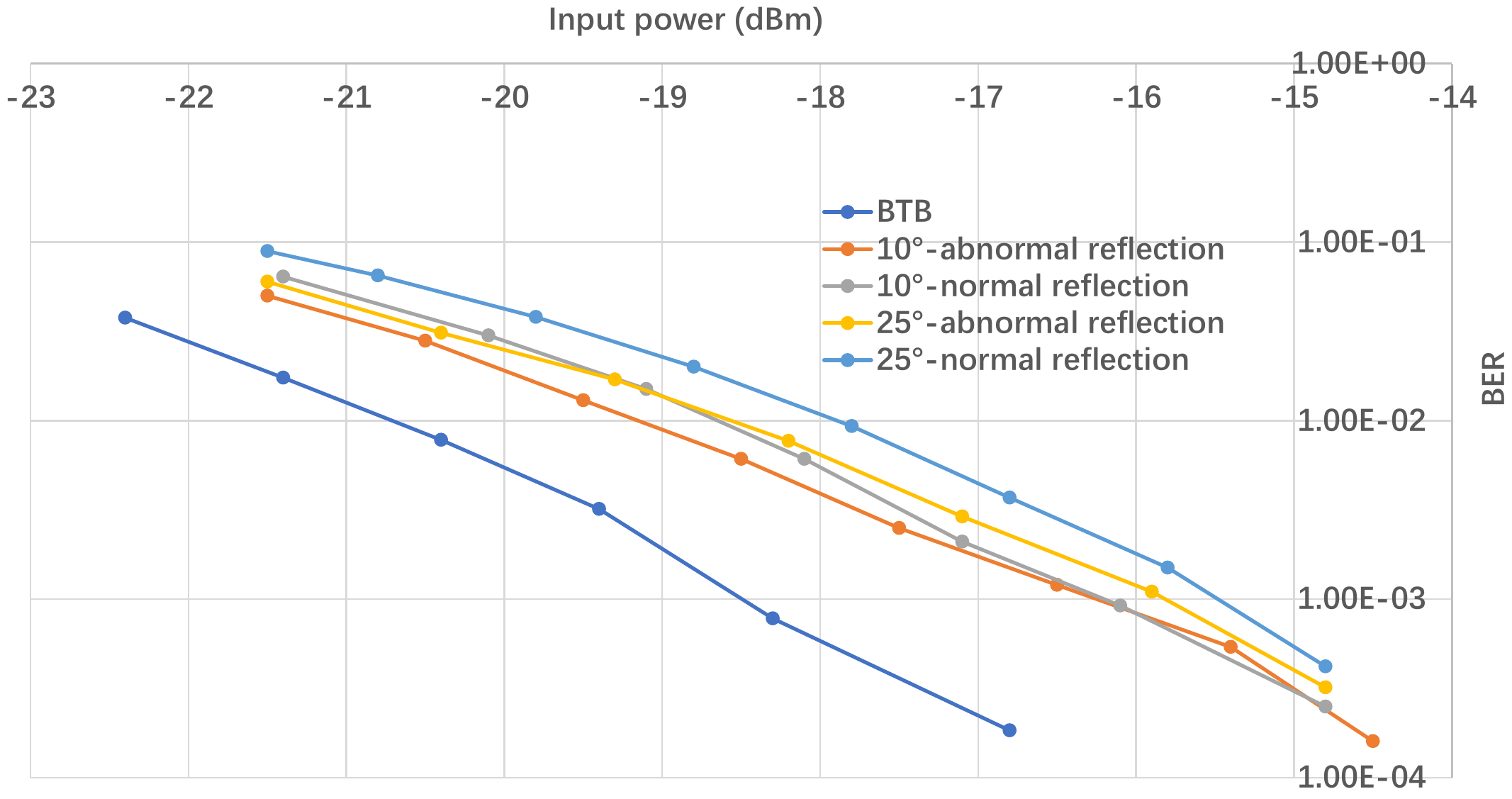}
	\caption{%
		BER performance of a 10-GBaud/s PAM4 signal over 1.2-m free space at different incident angles.
	}
	\label{fig12}
\end{figure*}

When $\theta_i>\theta_{ic}$, the abnormal reflection disappears, and surface plasmon-polaritons are excited. To obtain a broad incident range and meet the design requirement $\Lambda>2\lambda$, a large $\theta_{ic}$ is needed, resulting in a large $\Lambda$. Therefore, eight phase units are used to cover $2\pi$ ($\theta_{ic}=37.77^\circ$) with a $\pi/4$ phase interval. However, the $\pi/4$ phase interval cannot be achieved using only type-1 meta-atoms; therefore, type-2 meta-atom is introduced to achieve a large phase shift.

Figure 4c precisely shows the scattered phase of each phase unit, which uniformly covers a $2\pi$ phase. The simulation results agree very well with the model. \textbf{Figure 5} further shows the simulated $xy$-plane electric-field magnitude distribution in the center of the spacer of the designed meta-atoms. It can be seen that similar resonances occur in each meta-atom, and the resonance can be manipulated by varying the length of the top Au nano-pattern. Therefore, different phase responses can be obtained.

For an $x$-polarized incident plane wave, the phase response of each phase unit is basically the same. \textbf{Figure 6a} shows the simulated scattered $E_x$ phase patterns of all the phase units in a super cell under the illumination of normally incident $x$-polarized wave ($\lambda=1550$ nm). The corresponding phase values are shown in Figure 6b. Since two types of meta-atoms are used, a phase mismatch exists as we can see in Figure 6a. However, most of the phase units have the same phase response. Thus, in general, the super cell acts as a mirror.

Based on the above discussions, the designed metasurface has different reflection characteristics for $x$ and $y$-polarized incident light, that is, a polarization beam splitter. With this polarization beam splitter, one can manipulate the direction and power of the reflected light by controlling the polarization of the incident light.

A metasurface polarization beam splitter was fabricated in our clean room, as shown in \textbf{Figure 7}. The fabricated sample is measured at three wavelengths: 1530 nm, 1550 nm, and 1565 nm. \textbf{Figure 8} shows the measured relation between the incident angle and the abnormal reflection angle. These angle relations are measured in the $xz$-plane, as shown in Figure 3(a). It can be seen that the measurement results agree very well with the theory. The power efficiency of the abnormal reflection (normal reflection) is measured by adjusting the incident polarization to maximize the received abnormal reflection (normal reflection) power while keeping the incident power constant. At all the three wavelengths, the measured results show similar trends, as shown in \textbf{Figure 9}. For the abnormal reflection, the reflection efficiency decreases gently with increasing incident angle. The measured efficiency is approximately $80\%$ when the incident angle is less than $15^\circ$ at all the three wavelengths, and it is $>70\%$ when the incident angle is less than $25^\circ$, which is state-of-the-art efficiency. For the normal reflection, at all the three wavelengths, the measured reflection efficiency is $>80\%$ for most of the incident angles. There are two notches at $19^\circ$ and $26^\circ$, where the efficiency drops to approximately $70\%$ at 1550 nm and 1565 nm. At 1530 nm, the efficiency drops to approximately $60\%$. This is caused by the angular dispersion under TM-polarized illumination\cite{zhang2020controlling}. In general, the influence of the angular dispersion is limited, and a high reflection efficiency is achieved. The isolation between the normal and abnormal reflection can be measured in the same manner as the efficiency. As shown in \textbf{Figure 10}, the isolation shows the same trends as the efficiency. The overall isolation is $15\pm5$ dB for incident angles in the range of $10^\circ$ to $35^\circ$ at all the three wavelengths.

The measurement results show the designed metasurface polarization beam splitter has a large incident angle range and a high reflection efficiency. When the incident angle changes from $0^\circ$ to $35^\circ$, the abnormal reflected beam can cover a range of $50^\circ$, while the efficiency (for both normal and abnormal reflection) is larger than $50\%$ (3 dB). An incident angle of up to $35^\circ$ incident angle is available, which is close to the critical angle ($37.77^\circ$). The designed device also has a broad working bandwidth. The relation between the incident angle and the abnormal reflection angle remains almost unchanged when the input wavelength switches from 1530 nm to 1565 nm. Therefore, it can be used in the proposed 2D IR beam-steering system.

\section{System Demonstration}
\label{sec:Experiment}

A proof-of-concept experimental setup was built using the fabricated metasurface, as shown in \textbf{Figure 11}. A tunable laser source (TLS) with a 10-dBm output optical power is used to generate the optical carrier in the communication control center (CCC), which is then modulated by a 10-GHz Mach-Zehnder modulator (MZM). The electrical data are produced by an arbitrary waveform generator (AWG) to drive the MZM after amplification. In the experiment, transmitted pulse amplitude modulation (PAM) signals are generated offline using a MATLAB program and then sampled by the AWG running at 10 GSa/s, producing a 10-GBaud/s PAM-4 baseband signal. Thus, the achieved data rate is 20 Gbit/s. The received electrical PAM-4 signal is oversampled by the digital phosphor oscilloscope (DPO) sampling at 25 GSa/s. The modulated optical signal is amplified by an erbium-doped fiber amplifier (EDFA) and then launched into a 1.7-km single-mode fiber (SMF). A polarization controller (PC) combined with a free-space polarizer are used to control the polarization states of the incident light so as to manipulate the power distribution of beams emerging from the metasurface chip. The optical beam is launched into free space via a collimator (the focal length is 18 mm) with a measured power below 10 dBm which meets the requirements of human eye safety. The light is further collimated by a lens and then launched onto the chip, resulting in normal and abnormal beams. After 1.2-m free-space transmission, both beams are coupled into a short section of a multi-mode fiber and SMF via another collimator. The received optical power is amplified and then detected by an avalanche photodiode. The amplified output electrical signal is acquired by a DPO for further signal processing.

\textbf{Figure 12} illustrates the BER performance of the 20-Gbit/s PAM-4 signal as a function of the optical power received by a variable optical attenuator (VOA) placed at the front of the APD. The incident angles to the metasurface chip are set to $10^\circ$ and $25^\circ$, which correspond to a beam-steering range of $35^\circ$ (normal reflection $+$ abnormal reflection). Four output beams are generated, transmitted, and measured at the receiver end, as shown in Figure 12. We also measured the BER performance of the 20-Gbit/s PAM-4 signal at optical back-to-back (OBTB) transmission. Compared to OBTB transmission, a penalty of almost 3dB is observed at a $7\%$ forward error correction (FEC) limit of $1\times10^{-3}$, which is mainly caused by the noise figure of the EDFA at the receiver end.

\section{Conclusion}
\label{sec:Conclusion}

By combining active and passive approaches, we proposed a novel two-dimensional infrared beam-steering system, which can be tuned by changing the wavelength and polarization. The polarization tuning is achieved using a metasurface polarization beam splitter in conjunction with a liquid-crystal polarization controller. Grating loss is avoided by using metasurface, and high reflection efficiency is achieved. The wavelength tuning is enabled by a beam-steering module based on an arrayed waveguide grating router. The proposed system has scalability to support multiple beams, flexibility to steer the beam, high optical efficiency, and large coverage area. With the designed and fabricated metasurface polarization beam splitter, a 20-Gbps beam-steered infrared wireless link is built as proof of concept.

\section{Experimental Section}
\label{sec:Experimental Section}

\emph{Sample Fabrication}: The metasurface polarization beam splitter is fabricated using the following steps. Firstly, a 200-nm Au ground was deposited on a Si substrate by metal evaporation (BVR2008FC). Secondly, a 90-nm-thick SiO$_2$ layer was deposited on the Au ground by plasma-enhanced chemical vapor deposition (PECVD). Thirdly, a 950 PMMA A4 layer was spin-coated on the SiO$_2$ layer. After baking on a hotplate, the sample was sent into an electron beam direct write lithography system (EBPG5150) for pattern definition. After lithography, the sample was developed in an MIBK/IPA solution and then rinsed with IPA. Fourthly, a 2-nm-thick Cr layer was deposited on the sample by metal evaporation (BVR2008FC) to improve adhesion. Subsequently, a 53-nm-thick Au layer was deposited without taking the sample out. Finally, the top Au patterns were obtained by lift-off.

\begin{acknowledgement}
  J. Huang and C. Li contributed equally to the paper. This work is supported by the NWO Zwaartekracht program on Integrated Nanophotonics. The authors declare no conflict of interest.
\end{acknowledgement}

\bibliographystyle{lpr}
\bibliography{Reference3_5}

%
%
%
%
%
%
%
\end{document}